\documentclass{article}
\usepackage{mathptmx}
\usepackage{amsmath}
\usepackage{amssymb}
\usepackage{hyperref}
\usepackage{graphicx}
\usepackage{ulem}
\usepackage{setspace}
\hypersetup{pdfborder=false}
\title{An Invertible Linearization Map for the Quartic Oscillator}
\author{Robert L. Anderson{\footnote{\texttt{\href{mailto:andersonr@hal.physast.uga.edu}{andersonr@hal.physast.uga.edu}}}}\\
\vspace{.025in}\\
Department of Physics and Astronomy\\
\vspace{.025in}\\
University of Georgia\\
\vspace{.025in}\\
Athens, Georgia 30602}
\date{December 22, 2010}
\begin{document}                  

\maketitle

\begin{abstract}
{\noindent}The set of world lines for the non-relativistic quartic oscillator satisfying Newton’s equation of motion for all space and time in 1-1 dimensions with no constraints other than the ”spring” restoring force is shown to be equivalent (1-1-onto)
to the corresponding set for the harmonic oscillator. This is established via an energy preserving invertible linearization
map which consists of an explicit nonlinear algebraic deformation of coordinates and a nonlinear deformation of time
coordinates involving a quadrature. In the context stated, the map also explicitly solves Newton’s equation for the
quartic oscillator for arbitrary initial data on the real line. This map is extended to all attractive potentials given by
even powers of the space coordinate. It thus provides classes of new solutions to the initial value problem for all these
potentials. 
\end{abstract}

\part{INTRODUCTION}

The set of world lines for the non-relativistic quartic oscillator satisfying Newton’s equation of motion for all space and
time in 1-1 dimensions with no constraints other than the ”spring” restoring force is shown to be equivalent (1-1-onto)
to the corresponding set for the harmonic oscillator. This is established via an energy preserving invertible linearization
map which consists of an explicit nonlinear algebraic deformation of coordinates and a nonlinear deformation of time
coordinates involving a quadrature. In the context stated, this result also explicitly solves Newton’s equation for the
quartic oscillator for arbitrary initial data on the real line. No approximations are involved!\\

Specifically, each world line for the non-relativistic harmonic oscillator satisfying Newton’s equation of motion for all
space and time in 1-1 dimensions with no constraints other than the spring restoring force satisfies\\

\begin{align*}
m{\frac{d^2}{d{\hat{t}^2}}}{\hspace{4pt}}x{\hspace{4pt}}({\hat{t}})=-k_2{\hspace{4pt}}x{\hspace{4pt}}({\hat{t}})^2, \tag{1.1}    
\end{align*}

where as usual $m$=mass, $k_2$ = spring constant, and $k_2$ = $m{\omega}^2$ and is given by

\begin{align*}
x{\hspace{4pt}}({\hat{t}})={\sqrt{\frac{2E}{k_{2}}}}{\hspace{2pt}}cos{\hspace{2pt}}{\omega}{\hspace{4pt}}({\hat{t}}-{\hat{t}_{max}}),\tag{1.2}    
\end{align*}

for -${\infty}$ $<{\hat{t}}$ $<{\infty}$ and $E$ equals the total energy.

With the invertible map given in Part III, we shall pair each member of the set of all world lines for the harmonic oscillator with one for the quartic oscillator with the same mass and total energy obeying Newton's equation of motion:

\begin{align*}
m{\hspace{4pt}}{\frac{d^{2}}{{dt}^{2}}}{\hspace{4pt}}y{\hspace{4pt}}({t})=-{k_{4}}{\hspace{4pt}}y^{3}{\hspace{4pt}}(t).\tag{1.3}    
\end{align*}

The invertible map matches potential energies and momenta, hence energies.\\

It is important to comment that an extremal, i.e., each solution of the associated variational problem connecting the spacetime events $(x_{a}, {\hat{t}}_{a})$ and $(x_{b}, {\hat{t}}_{b})$ for the harmonic oscillator ( $(y_{a}, t_{a})$ and $(y_{b}, t_{b})$ for the quartic oscillator) is a segment of the world line with the same energy.  Hence, a fundamental role of our energy preserving linearization map is to map the set of extremals in spacetime connecting the initial point $(x_{a}, {\hat{t}}_{a})$ and the final point $(x_{b}, {\hat{t}}_{b})$ for the linear (harmonic) oscillator (ho), 1-1 onto the set of extremals in spacetime connecting the initial point $(y_{a}, t_{a})$ and the final point $(y_{b}, t_{b})$ for the quartic oscillator (qo).\\

R. C. Santos, J. Santos and J.A.S. Lima first demonstrated the possibility of linearization of the nonrelativisitic quartic oscillator to that of the nonrelativistic harmonic oscillator [1].  In [1], they approach the problem of dealing with nonlinear "oscillators" via the classical Hamilton-Jacobi equation in contrast to our approach via Newton's equations of motion.  The most important difference is we have to add a fundamental bookkeeping system to their work; namely, the signs of what we call $x$ and $y$ over a cycle have to be matched as well as requiring the harmonic oscillator time $\hat{t}$ and the quartic oscillator time $t$ to progress together in a positive manner.  Further, we match momenta which coupled with matching the potential energies allows us to match total energies.  (In [1], they match potential energies.)  This is critical since we not only map solutions of the quartic oscillator onto a solution of the harmonic oscillator, but this allows us to specify a 1-1 map of extremals in spacetime onto extremals in spacetime for the two systems.\\

In Part II it is shown that the world lines of any two harmonic oscillators can be mapped 1-1 onto each other.  So we can select any harmonic oscillator as representative in this context.  The essentials of the map we shall use in Part III to linearize the quartic oscillator are illustrated in the development in Part II where for this case both the space and time parts are algebraic.\\

In Part III, the set of world lines hence the spacetime extremals connecting the spacetime events ($y_{a}$, $t_{a}$) and ($y_{b}$, $t_{b}$) for the quartic oscillator particle system is shown to be equivalent to the set for the linear (harmonic) oscillator connecting the spacetime events ($x_{a}$, ${\hat{t}}_{a}$) and ($x_{b}$, ${\hat{t}}_{b}$) under an energy preserving map which is a nonlinear algebraic deformation of harmonic oscillator space coordinates and a nonlinear deformation of the harmonic oscillator time coordinates. The latter is given by quadrature and we shall deal with this point in Part III.  Part III ends with the ratio of the period of the quartic oscillator (${\tau}_{qo}$) to that of the harmonic oscillator period (${\tau}_{ho}$) which is shown to be inversely proportional to the fourth power of energy.\\

In Part IV, a summary of the extension of these results to the hierarchy of attractive potentials given by even powers of the space coordinate is given. With respect to [1], the remarks made in Part I generalize to potentials considered in Part IV.\\

In Part V, notable work in 1 + 1-Dim is briefly described and higher dimensional applications are referenced.\\

Finally in Part VI Concluding Remarks, three other possible applications are briefly described.

\part{EXTREMAL MAPPING $ho{\leftrightarrow}ho$}

The map to a second harmonic oscillator with mass $m$ and space coordinate ${\tilde{x}}$ is stated in two parts:\\

\paragraph{(A)}

\begin{align*}
{\tilde{x}}&={\sqrt{\frac{k_{2}}{\tilde{k_{2}}}}}{\hspace{4pt}}x\\
{\text{or}}&\\
x&={\sqrt{\frac{\tilde{k_{2}}}{k_{2}}}}{\hspace{4pt}}{\tilde{x}},\tag{2.1}\\
\end{align*}

This implements the physical requirement that ${\frac{1}{2}}k_{2}x^{2}$(${\hat{t}}$) = ${\frac{1}{2}}{\tilde{k}}_{2}{\tilde{x}}^{2}(\tilde{t})$ matching the potential energies at the two different times, coupled with matching of the signs of the space coordinates.\\

\paragraph{(B)}

\begin{align*}
{\frac{d{\tilde{t}}}{d{\hat{t}}}}&={\sqrt{\frac{k_{2}}{\tilde{k_{2}}}}},\\
{\text{and}}&\\
{\frac{d{\hat{t}}}{d{\tilde{t}}}}&={\sqrt{\frac{\tilde{k_{2}}}{k_{2}}}}.\tag{2.2}
\end{align*}

This follows by requiring\\

\begin{align*}
dx({\hat{t}})/d{\hat{t}}=d{\tilde{x}}({\tilde{t}})/d{\tilde{t}}.\tag{2.3}
\end{align*}

Given the matching of the potential energies, the matching of the velocities and the masses of the oscillators for all values of $k_{2}$ and ${\tilde{k}_{2}}$ implies physically matching the momentum and the kinetic energies at the two different times, i.e. $p_{ho}$(${\hat{t}}$) = ${\tilde{p}_{ho}}$(${\tilde{t}}$), $E_{ho}$ = ${\tilde{E}_{ho}}$ = $E$.\\

Equation (2.2) integrates to 

\begin{align}
{\tilde{t}}-{\tilde{t}_{max}}={\sqrt{\frac{k_{2}}{\tilde{k_{2}}}}}{\hspace{4pt}}({\hat{t}}-{\hat{t}_{max}})={\frac{\omega}{\tilde{\omega}}}({\hat{t}}-{\hat{t}_{max}})\tag{2.4}
\end{align}

Thus

\begin{align}
x({\hat{t}})={x_{max}}cos{\omega}({\hat{t}}-{\hat{t}_{max}}){\hspace{6pt}}{\Leftrightarrow}{\hspace{6pt}}{\tilde{x}}({\tilde{t}})={\tilde{x}_{max}}cos{\tilde{\omega}}({\tilde{t}}-{\tilde{t}_{max}})\tag{2.5}
\end{align}

or this map implements the correspondence between Newton's equations of motion\\

\begin{align*}
m{\frac{d^{2}}{d^{2}{\hat{t}}}}x({\hat{t}})=-k_{2}x({\hat{t}}){\hspace{6pt}}{\Leftrightarrow}{\hspace{6pt}}m{\frac{d^{2}}{d^{2}{\tilde{t}}}}{\tilde{x}}({\tilde{t}})=-{\tilde{k}}_{2}{\tilde{x}}({\tilde{t}}).\tag{2.6}
\end{align*}

Finally, all the world lines of the hat system correspond to all the world lines of the tilde system, so we are free to pick any harmonic oscillator as a representative.  We designate the hat system as our choice.

\part{EXTREMAL MAPPING $ho{\leftrightarrow}qo$}

The map to the quartic oscillator with mass $m$ and space coordinate $y$ is stated in two parts:\\

\paragraph{(A)}

\begin{align*}
y&=(2k_{2}/k_{4})^{1/4}x/(x^{2})^{1/4}\\
{\text{or}}&\\
x&=(k_{4}/2k_{2})^{1/2}(y^{2})^{1/2}y,\tag{3.1}
\end{align*}

where $y$ is the space coordinate of the quartic oscillator and we have used the representation sgn($x$) = $x$/($x^{2}$)$^{1/2}$ and similarly for sgn($y$).  This implements the physical requirement that ${\frac{1}{2}}k_{2}x^{2}$(${\hat{t}}$) = ${\frac{1}{4}}k_{4}y^{4}$($t$) i.e. matching the potential energies at the two different times, coupled with matching of the signs of the space coordinates.\\

\paragraph{(B)}

\begin{align*}
{\frac{dt}{d{\hat{t}}}}&=1/2(2k_{2}/k_{4})^{1/4}(x^{2}({\hat{t}}))^{-1/4},\\
{\text{and}}&\\
{\frac{d{\hat{t}}}{dt}}&=(2k_{4}/k_{2})^{1/2}(y^{2}(t))^{1/2},\tag{3.2}
\end{align*}

which results by requiring\\

\begin{align*}
dx({\hat{t}})/d{\hat{t}}=dy(t)/dt.\tag{3.3}
\end{align*}

Given the matching of the potential energies, the matching of the velocities and the masses of the oscillators for all values of $k_{2}$ and $k_{4}$ implies physically matching the momentum and the kinetic energies at the two different times, i.e. $p_{ho}$(${\hat{t}}$) = $p_{qo}$($t$), $E_{ho}$ = $E_{qo}$ = $E$.  It is interesting to note that (2.2) is also equivalent to requiring $dt$/$d{\hat{t}}$ = [($-k_{4}y^{3}$($t$))/($-k_{2}x$(${\hat{t}}$))]$^{-1}$ = the inverse of the ratio of the forces.\\

This map implements the correspondence between Newton's equations of motion\\

\begin{align*}
m{\frac{d^{2}}{d^{2}{\hat{t}}}}x({\hat{t}})=-k_{2}x({\hat{t}}){\hspace{6pt}}{\Leftrightarrow}{\hspace{6pt}}m{\frac{d^{2}}{d^{2}t}}y(t)=-k_{4}y^{3}(t)\tag{3.4}
\end{align*}

in the following straight-forward way.  Differentiation of (2.1) shows that (2.1) and (2.2) are self-consistent.  Differentiating (2.1) a second time, invoking (2.2) and assuming the validity of one side of (2.6) then yields the validity of the other side of the equivalence in (2.6).\\

If one plots the $ho$-potential and the $qo$-potentials vs space coordinates, then the  horizontal lines in such a plot allow one to graphically read off the corresponding coordinates.   Because of reflection symmetry in the vertical axis, one need only plot the potential for the positive space coordinate.  Further, for some $y$-coordinate near enough to zero but not zero, the $qo$-potential is less than the $ho$-potential for all nonzero $y$-coordinates less than that value. This feature is illustrated in Figure 1.\\
\\

\begin{figure}[h!]
\includegraphics{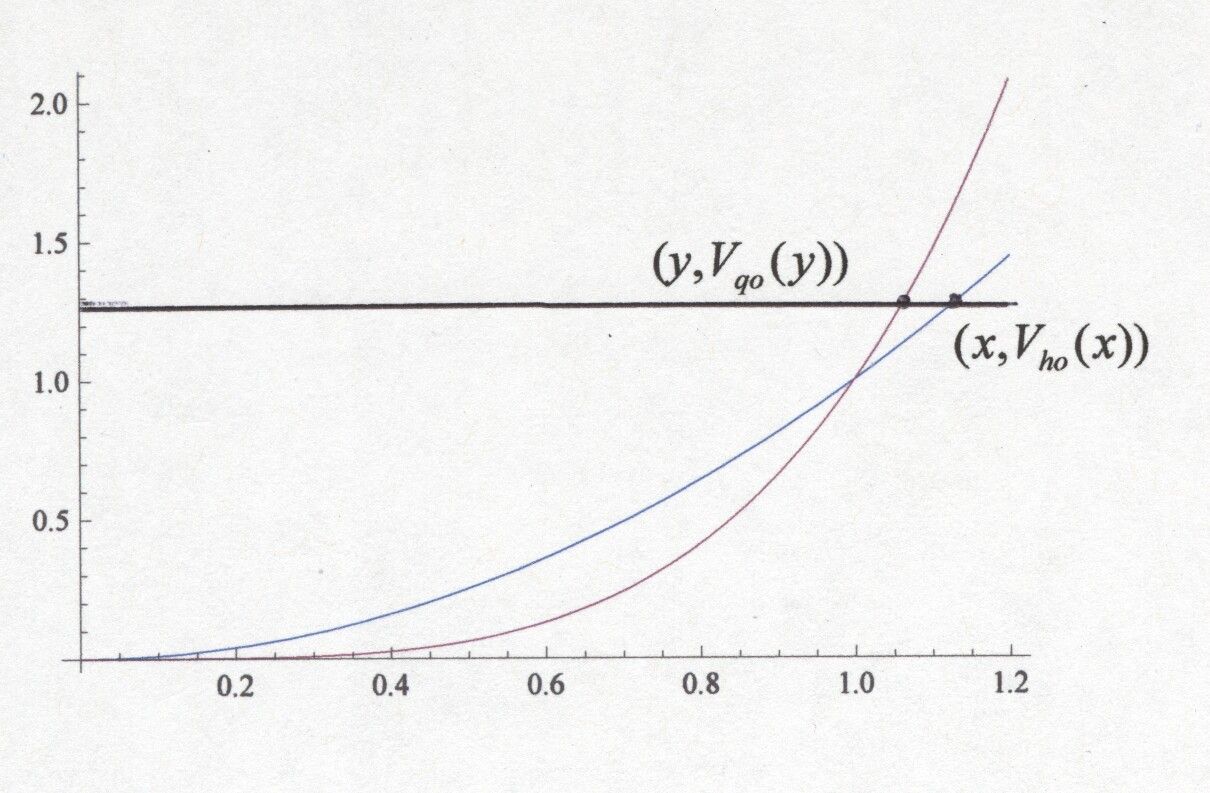}
\centering
\caption{Comparison of Potentials}
\centering
\end{figure}

In [1], they approach the problem of dealing with nonlinear ``oscillators" via the classical Hamilton-Jacobi equation in contrast to our approach via Newton's equations of motion. Another way that our work differs from theirs is that we do not identify the solutions of the HJ equation, rather we only assume their existence.  Another difference is we have to add a fundamental bookkeeping addition to their work; namely, the signs of what we call $x$ and $y$ over a cycle  have to be matched as well as requiring the harmonic oscillator time ${\hat{t}}$ and the quartic oscillator time $t$ to progress together in a positive manner. Further, we match momenta which, coupled with matching the potential energies, allows us to match total energies. (In [1], they match potential energies.) This is critical since we not only map solutions of the quartic oscillator onto a solution of the harmonic oscillator, this allows us to specify a 1-1 map of world lines in spacetime onto world lines in spacetime for the two systems.\\

\begin{figure}[h!]
\includegraphics{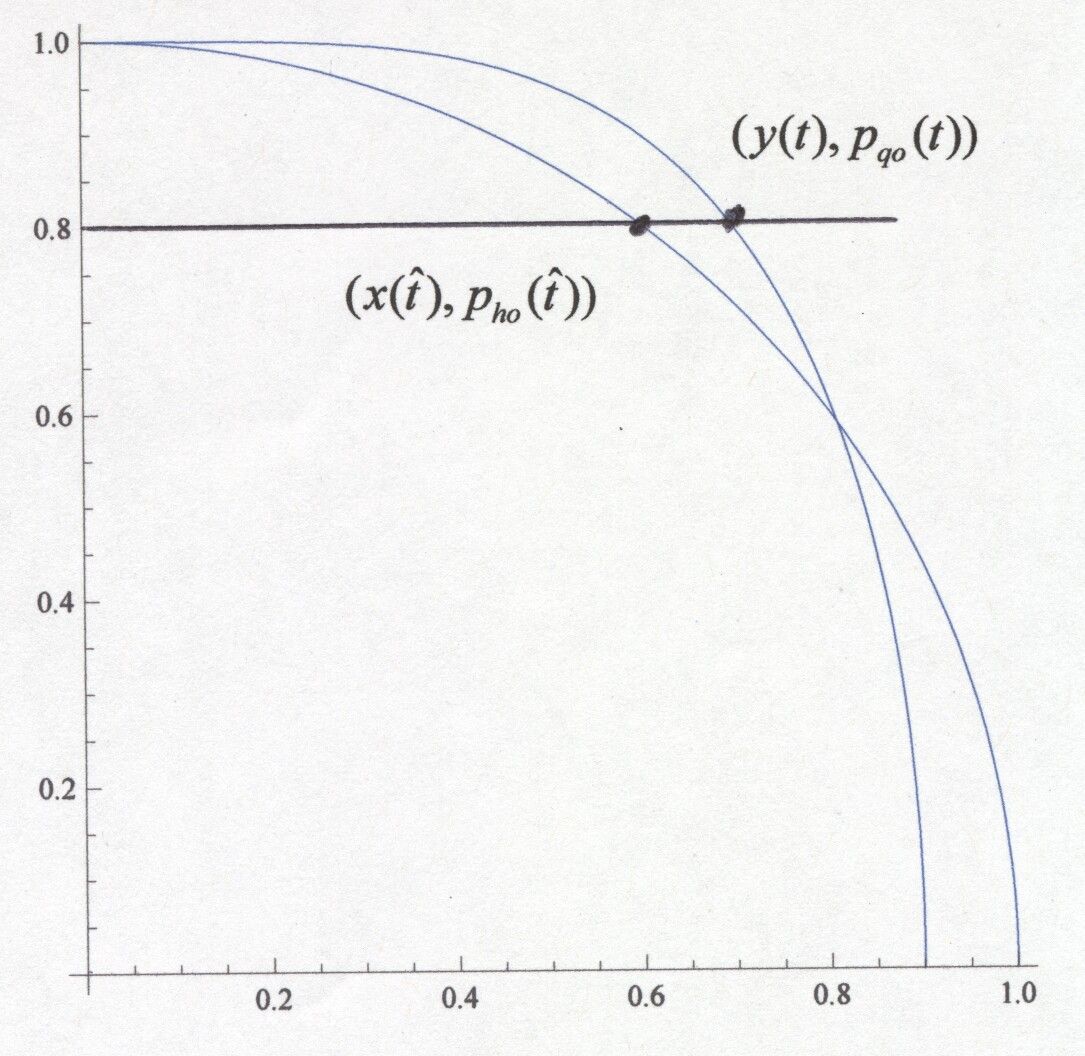}
\centering
\caption{Phase Space}
\centering
\end{figure}

\vspace{4in}

The corresponding phase space diagrams are interesting, but very case dependent in terms of $k_{4}$/$k_{2}$ and $E_{ho}$.  The horizontal coordinate scales are set by $x_{\max}$ = ${\sqrt{2E/k_{2}}}$ and $y_{\max}$ = $\sqrt[4]{4E/k_{4}}$.  As a result $y_{\max}$ can be less than, greater than or equal to $x_{\max}$.  The vertical scales are set by the equality of the linear momentum at the different times, i.e. $m{\frac{dy}{dt}}$($t$) = $m{\frac{d}{d{\hat{t}}}}x$(${\hat{t}}$) which also implies that $m{\frac{dy}{dt}}{\biggl{\vert}}_{\max}$ = $m{\frac{d}{d{\hat{t}}}}x{\biggl{\vert}}_{\max}$.  If one were to draw a phase diagram, horizontal lines yield the corresponding phase space coordinates ($x$(${\hat{t}}$), ${\frac{dx}{d{\hat{t}}}}$(${\hat{t}}$)) $\Leftrightarrow$ ($y$($t$), ${\frac{dy}{dt}}$($t$)).  Thus, while both make one complete cycle together in terms of phase coordinates, they do it in different times. This point is important to emphasize; one complete cycle for the $ho$ in the phase space diagram corresponds to one complete cycle for the $qo$ in the phase space diagram, but this is done in general in different periods.  These features are illustrated in Figure 2 for $y_{\max}$ $<$ $x_{\max}$.  Because of symmetry, one need only graph the first quadrant of the phase diagram.\\

Now, the nonlinear deformation of time given by (3.2) is in terms of a definite integral.
\begin{align*}
(1/2){\hspace{6pt}}(2k_{2}/k_{4})^{1/4}{\hspace{6pt}}{\int^{{\hat{t}}_{b}}_{\hat{t}_{a}}}{\hspace{6pt}}{\frac{1}{{\biggl{(}}(x^{2}({\hat{t}}))^{1/4}{\biggr{)}}}}{\hspace{6pt}}d{\hat{t}}=t_{b}-t_{a}\tag{3.5}
\end{align*}

Equation (3.5) is the operative quadrature that establishes the connection between the two ``times" $t$ and ${\hat{t}}$.  Mathematica [2] provides a complete program for evaluating this integral.\\

Now we turn to evaluating (3.5) for the period of the quartic oscillator ${\tau_{qo}}$ in terms of the period ${\tau_{ho}}$ of the harmonic oscillator.  It is sufficient and convenient to set  ${\hat{t}_{max}}$ = ${\pi}/2$ in (1.2) or take the world line corresponding to

\begin{align*}
x({\hat{t}})=x_{\max}{\sin}{\omega}{\hat{t}},\tag{3.6}
\end{align*}

where as before $x_{\max}$ = ${\sqrt{(2E/k_{2})}}$.  Then (3.5) becomes for one cycle of the quartic harmonic oscillator (or linear oscillator)\\
\begin{align*}
1/2{\hspace{6pt}}(k^{2}_{2}/k_{4}E)^{1/4}{\hspace{6pt}}{\int^{{\tau}_{ho}}_{0}}{\hspace{6pt}}{\frac{1}{{\biggl{(}}({\sin}^{2}{\omega}{\hat{t}})^{1/4}{\biggr{)}}}}{\hspace{6pt}}d{\hat{t}}={\tau}_{qo}.\tag{3.7}
\end{align*}

Equation (3.7) is an elliptic function of the first kind [2].\\

A simplification is possible, namely the change of variable $\theta$ = ${\omega}{\hat{t}}$.  Equation (3.7) then becomes\\

\begin{align*}
1/2{\hspace{6pt}}{\biggl{(}}k^{2}_{2}/k_{4}E{\biggr{)}}^{1/4}{\hspace{6pt}}{\omega}^{-1}{\hspace{6pt}}{\int^{2{\pi}}_{0}}{\hspace{6pt}}{\frac{1}{{\biggl{(}}({\sin}^{2}{\theta})^{1/4}{\biggr{)}}}}{\hspace{6pt}}d{\theta}={\tau}_{qo},\tag{3.8}
\end{align*}

where, again, $\omega$ = 2$\pi$/${\tau}_{ho}$.\\

Equation (3.8), using symmetry, yields the period ${\tau}_{qo}$ of the oscillator in terms of  the period ${\tau}_{ho}$ as follows\\
\begin{align*}
{\tau}_{qo}/{\tau}_{ho}=(k^{2}_{2}/k_{4}E)^{1/4}{\hspace{6pt}}[{\pi}^{-1}{\hspace{6pt}}{\int^{{\pi}/2}_{0}}{\hspace{6pt}}({\sin}{\theta})^{-1/2}{\hspace{6pt}}d{\theta}],\tag{3.9}
\end{align*}

where using Mathematica [2] the term in square brackets equals .83.  It follows from (3.8) that for fixed $k_{2}$ and $k_{4}$ we have ${\tau}_{qo}$/${\tau}_{ho}$ $\approx$ $E_{^{\mbox{.}}}^{-1/4}$\\

\part{EXTREMAL MAPPING FOR ${\frac{1}{2n}}$ $k_{2n}y_{2n}^{2n}$($t$) $|_{n>1}$ HIERARCHY}

In this section we present a summary of the extension of these results to the hierarchy of attractive potentials given by even powers of the space coordinate paralleling that given in Part II - Part III.\\

First, we shall outline the mapping of the harmonic oscillator extremals onto the extremals  of a each member of an hierarchy of  attractive oscillators with coordinates $y_{2n}$($t$); $n$ = 2, 3, 4, ... characterized by even positive power law potentials. (The case, $y$ $\equiv$ $y_{4}$ which is included in the hierarchy, has been the subject of the preceding paragraph.)  In a straight forward manner the mappings in Part III, generalize and yield the following relationships: \\

\paragraph{(A)}

\begin{align*}
y_{2n}&=(nk_{2}/k_{2n})^{1/2n}x(x^{2})^{(1/2)(1-n)/n}\\
x&=(k_{2n}/nk_{2})^{1/2}y_{2n}(y^{2}_{2n})^{(n-1)/2},\tag{4.1}
\end{align*}

which is the generalization of (3.1).  The generalization of (3.2) is given by:\\

\paragraph{(B)}

\begin{align*}
{\frac{dt}{d{\hat{t}}}}&=n^{-(2n-1)/2n}(k_{2}/k_{2n})^{1/2n}(x^{2}({\hat{t}}))^{-(n-1)/2n}({\hat{t}}),\\
{\mbox{and}}\\
{\frac{d{\hat{t}}}{dt}}&={\sqrt{n}}(k_{2n}/k_{2})^{1/2}(y^{2}_{2n})^{(n-1)/2}.\tag{4.2}
\end{align*}

These mappings take the space-time extremals of the linear oscillator with coordinates ($x$, ${\hat{t}}$) and map them onto the space-time extremals of the 2$n${\textsuperscript{th} oscillator with coordinates ($y_{2n}$, $t$).\\

A straightforward calculation yields\\

\begin{align*}
m{\frac{d^{2}}{d^{2}{\hat{t}}}}x({\hat{t}})=-k_{2}x({\hat{t}}{\hspace{4pt}}{\Leftrightarrow}{\hspace{4pt}}m{\frac{d^{2}}{d^{2}t}}y_{2n}(t)=-k_{2n}y_{2n}^{2n-1}(t).\tag{4.3}
\end{align*}

Further, as a consequence of the above, we have the following equality for the conserved total energies\\

\begin{align*}
E_{2}={\frac{1}{2}}m({\frac{dx_{2}}{d{\tau}}})^{2}{\hspace{4pt}}+{\hspace{4pt}}{\frac{1}{2}}k_{2}x^{2}_{2}=E_{2n}={\frac{1}{2}}m({\frac{dx_{2n}}{dt}})^{2}{\hspace{4pt}}+{\hspace{4pt}}{\frac{k_{2n}}{2n}}q_{2n}^{2n}=E,\tag{4.4}
\end{align*}

The operative deformation of time given by (4.2) becomes in integral form

\begin{align*}
t_{b}-t_{a}={\int^{{\hat{t}}_{b}}_{{\hat{t}}_{a}}}n^{-(2n-1)/2n}(k_{2}/k_{2n})^{1/2n}(x^{2}({\hat{t}}))^{-(n-1)/2n}d{\hat{t}}\tag{4.5}
\end{align*}

This is implemented using Mathematica [2].

All of the analyses presented in Part III can then be extended to the members of the hierarchy.\\

\part{NOTABLE CONSERVATIVE 1 + 1-DIM WORK}

Whittaker [3, p. 64] solved all conservative 1 + 1-Dim mechanical systems up to a quadrature by solving the energy relationship for the velocity and integrating:

\begin{align*}
t_{b}-t_{a}={\int^{y_{b}}_{y_{a}}}{\pm}{\sqrt{2m}}{\sqrt{E-V(y')}}dy'.\tag{5.1}
\end{align*}

For the quartic oscillator, we can cast (5.1) into the following form

\begin{align*}
t_{b}-t_{a}={\int^{{\theta}_{b}}_{{\theta}_{a}}}{\frac{1}{2{\sqrt{2mE}}(sin^2{\theta})^{1/2}}}d{\theta},\tag{5.2}
\end{align*}

where

\begin{align*}
{\sqrt{\frac{k_4}{4E}}}(y^2)^{1/2}y=sin{\theta}.\tag{5.3}
\end{align*}

We have accounted for the sign of the velocity with $cos{\theta}$.\\

Similar forms can be derived for the rest of the attractive hierarchy studied in this paper.\\

One of the most interesting formulations is the one found by a "regularization" scheme for the differential equations [4, p. 14-17] describing the negative energy 1 + 1-Dim Kepler problem. Interestingly, this leads to a cycloid solution and related harmonic oscillator solution. This problem has an old and distinguished past and the author has not done justice to it here. This solution is not obtainable by matching of potential and kinetic energies as employed in this paper. The utility of the 1 + 2-Dim and the 1 + 3-Dim versions of their regularization in the Feynman Path Integral Method is prominent in the work of [5], [6].\\

The radical solutions of 1 + 2-Dim central force problems with angular momentum as one constant of the motion and energy as the other are addressed by Whittaker [3, p. 80-81]. He solves them in the spirit of the 1 + 1-Dim problem cited above, but in two dimensions it requires two quadratures.\\

\part{CONCLUDING REMARKS}

Physically what we have done with our linearization map is to view the tape of the space-time evolution of these nonlinear systems using the right optical lens to accomplish the nonlinear deformation of space and playing each frame according to a nonlinear deformation of time.  When we do this we see that the quartic oscillator evolves like an harmonic oscillator.\\

The mappings given in Parts III-IV provide new classes of exact solutions for nonlinear spring systems.\\

The essence of the maps presented here is that they involved two conservative mechanical systems and if it makes sense they match the potential energies and momenta of the two systems.  Within this context, its utility arises if solutions (a solution) of one of the systems are (is) known.  This means e.g., that one could linearize the hierarchy

\begin{align*}
-{\frac{K_{2n}}{2n}}{z^{2n}}{\vert_{n>1}}{\hspace{4pt}}to-{\frac{K}{2}}{z^{2}},
\end{align*}

or one could treat each motion of a quartic oscillator bouncing off an infinite barrier as a sequence of maps to a harmonic oscillator capturing the discontinuous velocity at the barrier, or one could map the radial equation in the procession of Mercury onto that of the inverse law force (4), [7, p. 194].  The latter involves extracting the zeros of polynomials cubic in ${\mu}={\frac{1}{{r}}}$ for the algebraic part of the map.  This is a nonlinear problem to nonlinear problem in the context of boundstates.\\

Special thanks go to my Department of Mathematics colleague Robert Varley.  He spent enumerable hours over a four year period of time discussing this work with me.  His comments, questions and posing of challenging related problems helped to clarify for me many aspects of this work.\\

The author wishes to aknowledge the referee's role in particular, in directing him to the work in [5] and [6].\\

The author wishes to thank Professor Howard Lee for insightful discussions and his constant encouragement. The idea to emphasize the quartic oscillator was his.  \\

\section*{References}

[1] R. C. Santos, J. Santos and J. A. S. Lima, ``Hamilton-Jacobi Approach for Power-Law Potentials", Braz. J. Phys. {\textbf{36}}  04.A, (2006) pp. 1257-1261.

[2] Stephen Wolfram, {\textit{Mathematica Version 7.0}}.

[3] E. T. Whittaker, {\textit{A Treatise on the Analytical Dynamics of Particles and Rigid Bodies}}, 3rd ed. (University of Cambridge Press, 1927).

[4] E. L. Stiefel and G. Scheifele, {\textit{Linear and Regular Celestial Mechanics}}, (Berlin/Heidelberg/New York 1971. Springer-Verlag 1971) Chapter 1.

[5] P. Kustaanheimo and E. Stiefel, ``Perturbation Theory of Kepler Motion Based on Spinor Regularization", (J. Reine Angew. Math 218, 204-219, 1965).

[6] I.H. Duru and H. Kleinert, ``Quantum Mechanics of H-Atom from Path Integrals", (Fortschr. Physik 30, 401-435, 1982).

[7] James B. Hartle, {\textit{Gravity/An Introduction to Einstein's General Relativity}}, (Addison Wesley 2003).  

\end{document}